\documentclass[conf]{new-aiaa}
\usepackage[utf8]{inputenc}

\usepackage{graphicx}
\usepackage{amsmath}
\usepackage[version=4]{mhchem}
\usepackage{siunitx}
\usepackage{longtable,tabularx}
\usepackage{subcaption} %
\usepackage{booktabs}
\usepackage{float}
\usepackage{verbatim}

\usepackage{natbib} %
\hypersetup{hidelinks}

\title{DSMC Study of Hypersonic Boundary-Layer Unsteadiness Using Data-Driven Analysis}
\author{Mert Senkardesler\footnote{Graduate Student, Department of Aerospace Engineering, AIAA Student Member.}, Irmak T. Karpuzcu\footnote{Graduate Student, Department of Aerospace Engineering, AIAA Student Member.} and Deborah A. Levin\footnote{Professor, Department of Aerospace Engineering, AIAA Fellow}}
\affil{University of Illinois, Urbana-Champaign, IL, USA}
\author{Vassilis Theofilis\footnote{Professor, The Steven B Kline Department of Aerospace Engineering, AIAA Associate Fellow}}
\affil{Technion -- Israel Institute of Technology, Haifa, Israel}
\date{}

\begin{document}

\maketitle
\begingroup\renewcommand{\thefootnote}{}\footnotetext{Presented at the AIAA AVIATION 2026 Forum, 8-12 June 2026, San Diego, CA. Published by the American Institute of Aeronautics and Astronautics, Inc. Cite as: M. Senkardesler, I. T. Karpuzcu, D. A. Levin, and V. Theofilis, ``DSMC Study of Hypersonic Boundary-Layer Unsteadiness Using Data-Driven Analysis,'' AIAA Paper 2026-4737, AIAA AVIATION 2026 Forum, San Diego, CA, June 2026, \url{https://doi.org/10.2514/6.2026-4737}. Copyright \textcopyright{} 2026 by the American Institute of Aeronautics and Astronautics, Inc. All rights reserved. This is the authors' submitted version, posted in accordance with the AIAA self-archiving and posting policy.}\addtocounter{footnote}{-1}\endgroup

\begin{abstract}
The prediction of transition remains challenging because the receptivity amplitude is usually prescribed rather than measured. This work uses Direct Simulation Monte Carlo (DSMC) to study an unforced Mach 6 nitrogen boundary layer over a 120 mm flat plate, with no imposed inlet wave, wall signal, modal shape, or disturbance spectrum. The retained unsteady field is the intrinsic finite-particle stochastic fluctuation field carried by the DSMC representation. Changing the DSMC particle count provides an independent numerical control parameter that sets a predictable finite-particle noise floor while preserving the time-averaged boundary-layer flow. Dynamic Mode Decomposition (DMD) is then applied to unforced DSMC snapshots. The DMD eigenvalues and spatial modes estimate the quantities that appear in the LST normal-mode description, including frequency, temporal growth or decay, wavenumber, wavelength, phase speed, mode location, and amplification envelope. LST is used after extraction to test whether the data-derived structures occupy the expected second-mode band. The results show that DSMC data can connect a measured finite-particle stochastic disturbance level to second-mode instability dynamics without imposed disturbance spectra.
    
\end{abstract}

\section{Introduction}

Hypersonic boundary-layer transition depends on the disturbance field entering the boundary layer and on the downstream amplification of its unstable components. For the Mach 6 flat-plate flow considered here, the relevant high-frequency instability is Mack's second mode. Linear stability theory, direct numerical simulation, and experiments provide a well-developed description of its unstable frequencies, growth regions, wavelengths, and phase speeds once the input disturbance is specified~\cite{mackBoundaryLayerLinearStability,kennethfstetsonHypersonicBoundaryLayer1980,frankoDirectNumericalSimulation2011,zhongBoundarylayerReceptivityMach2006}. The less constrained quantity is the receptivity input: the amplitude, frequency content, and spatial form of the disturbances that reach the boundary layer. Freestream acoustic noise, entropy fluctuations, vortical disturbances, wall motion, roughness, and particulates can all provide input disturbances, but their amplitudes are case dependent~\cite{schneiderHypersonicLaminarTurbulent2004}. In low-disturbance or high-altitude environments, and near leading edges where externally imposed disturbances may be weak, intrinsic molecular fluctuations can become a measurable part of that input.

Direct Simulation Monte Carlo (DSMC) is useful for this problem because the gas is represented by molecular motion and collision rather than by a deterministic continuum field. A pressure, density, temperature, or velocity sampled in a finite volume is therefore a finite molecular sample, and its fluctuation level depends on the number of independent molecular states represented in that sample. In DSMC, the computational particle weight, namely the number of real molecules represented by each simulation particle, also controls the finite-particle sampling level. Following the thermal-fluctuation receptivity picture of Fedorov and Tumin~\cite{fedorovReceptivityHighSpeedBoundary2017}, the present work treats the DSMC fluctuation field as a measurable stochastic input, while keeping its dependence on the chosen particle weight explicit.

To separate finite-particle sampling effects from changes in the mean state, two DSMC simulations are compared at the same Mach number, wall temperature, domain, and sampling interval, but with different computational particle weights and total simulation particle counts. The mean boundary-layer profiles remain unchanged, while the RMS fluctuation level follows the expected particle-weight scaling; the RMS change is therefore attributed to finite-particle sampling rather than to a changed mean flow. This comparison defines a reproducible disturbance-amplitude scale for interpreting DMD modes extracted from the unforced DSMC fields. Once that fluctuation level is characterized, the instability analysis is posed directly from the stored unsteady fields. The DSMC snapshots contain the boundary-layer response to intrinsic stochastic fluctuations, not the response to a selected inlet wave, wall signal, or assumed instability shape. Dynamic Mode Decomposition (DMD) then extracts spatio-temporal modes from those snapshots~\cite{SCHMID_2010}. For convective second-mode waves, the DMD eigenvalues provide the modal frequency and temporal growth or decay, while the spatial phase and amplitude of each mode provide the streamwise wavenumber, wavelength, phase speed, mode location, amplitude envelope, and local amplification or decay. Thus DMD converts unforced DSMC macrofield snapshots into data-derived instability characteristics.

LST is used here as an a posteriori technique. The Mach 6 flat-plate second-mode problem is used because it has a known stability reference: LST predicts the expected unstable frequency band, growth region, modal wavelength, phase speed, and downstream location of the waves. The present question is whether those same instability characteristics can be obtained from unforced DSMC data without using LST neutral curves, prescribed mode families, external seeding, or assumed disturbance spectra as inputs. Agreement with LST then checks whether the extracted structures match second-mode waves. The same DSMC-DMD procedure can later be applied where no reliable modal prior exists, including rarefied leading-edge regions, shock-layer and entropy-layer interactions, bluntness effects, wall-motion coupling, and other nonequilibrium hypersonic flows where continuum stability assumptions are incomplete.

\section{Numerical Methods and Flow Conditions}

This work models a Mach 6 nitrogen flow over a flat plate using the Direct Simulation Monte Carlo (DSMC) method~\cite{birdMolecularGasDynamics1994}. DSMC is a kinetic solver of the Boltzmann equation in which the gas is represented by computational particles whose motion and stochastic collisions recover the kinetic evolution of the flow. The stored DSMC macrofields include finite-particle stochastic fluctuations, which provide the disturbance field analyzed below. The primary simulation is the 5B particle-count 120 mm short-plate continuation case, where 5B denotes a total simulation particle count of approximately \(5.0\times10^9\). Intermolecular collisions are modeled with the Variable Hard Sphere (VHS) model~\cite{birdMolecularGasDynamics1994} and the majorant frequency scheme~\cite{nanbuMajorantFrequencyScheme1997}. The study uses the SMILE-DSMC solver~\cite{ivanovStatisticalSimulationMonte2012}. The essential freestream, numerical, and sampling parameters are summarized in Table~\ref{tab:DSMCFLOW}.

\begingroup
\setstretch{1.0}
\begin{table}[H]
\centering
\small
\setlength{\tabcolsep}{3pt}
\caption{Freestream values and DSMC numerical parameters for the primary 5B case.}
\label{tab:DSMCFLOW}
\begin{minipage}[t]{0.43\textwidth}
\begin{tabular}[t]{@{}p{0.62\linewidth}p{0.34\linewidth}@{}}
\toprule
\multicolumn{2}{@{}l}{\textbf{Freestream Conditions}} \\
\midrule
Working gas                         & Nitrogen \\
Mach number \(M_\infty\)            & 6 \\
Velocity \(U_\infty\)               & 860.6 m/s \\
Temperature \(T_\infty\)            & 50 K \\
Pressure \(p_\infty\)               & 828.3 Pa \\
Density \(\rho_\infty\)             & \(5.58\times10^{-2}\) kg/m\(^3\) \\
Number density \(N_\infty\)         & \(1.2\times10^{24}\) m\(^{-3}\) \\
Mean free path \(\lambda_\infty\)   & \(6.148\times10^{-7}\) m \\
\(Re_\infty/L\)                     & \(1.176\times10^7\) m\(^{-1}\) \\
Wall temperature \(T_w\)            & 300 K \\
\bottomrule
\end{tabular}
\end{minipage}
\hfill
\begin{minipage}[t]{0.52\textwidth}
\begin{tabular}[t]{@{}p{0.52\linewidth}p{0.44\linewidth}@{}}
\toprule
\multicolumn{2}{@{}l}{\textbf{DSMC Numerical and Sampling Parameters}} \\
\midrule
Domain                             & \(x=0\)--120 mm, \(y=0\)--3 mm \\
Sampling grid                      & \(2402\times60\) \\
Time step \(\Delta t\)             & 1.0 ns \\
Stored-snapshot period             & 50 steps, 50 ns \\
Analysis samples               & 1,001 \\
Particle weight \(FNUM\)           & \(8.525\times10^{10}\) \\
Total simulation particles \(N_p\) & \(\approx5.0\times10^9\) \\
Particles per \(\lambda_\infty^2\) & 5.32 \\
Flow species                     & Nitrogen \(\mathrm{N}_2\) \\
Production resources               & 80 nodes, 3840 MPI ranks \\
\bottomrule
\end{tabular}
\end{minipage}
\end{table}
\endgroup

The VHS viscosity law is \(\mu=\mu_{\mathrm{ref}}(T/T_{\mathrm{ref}})^\sigma\), with \(\mu_{\mathrm{ref}}=1.458731\times10^{-5}\) Pa\,s at \(T_{\mathrm{ref}}=273.15\) K and \(\sigma=0.75\). The stored scalar fields are \(n,u,v,T,p\) on the \(60\times2402\) sampling grid.

\begingroup
\setstretch{1.0}
\begin{table}[H]
\centering
\small
\caption{Particle-count cases used for the RMS comparison and mean-flow resolution check. The table lists the total simulation particle count and the freestream particles per \(\lambda_\infty^2\).}
\label{tab:particlecountcompare}
\begin{tabular}{@{}lccc@{}}
\toprule
Case & \(FNUM\) & \(N_p\) & Particles per \(\lambda_\infty^2\)\\
\midrule
Coarse DSMC & \(2.88\times10^{11}\) & \(\approx1.5\times10^9\) & 1.58 \\
5B & \(8.525\times10^{10}\) & \(\approx5.0\times10^9\) & 5.32 \\
\bottomrule
\end{tabular}
\end{table}
\endgroup

\noindent Table~\ref{tab:particlecountcompare} isolates the  case-to-case numerical changes used in the RMS study: particle weight, total simulation particle count, and particles per \(\lambda_\infty^2\). The coarse DSMC case is included as a deliberately lower-particle-count control. Its role is to check whether the mean boundary layer is already insensitive to particle count and has therefore converged to the correct flow state. For both particle-count cases, the mean-profile and RMS comparisons use stored timesteps 650,000--700,000, giving 1,001 samples separated by \(50\) ns. Table~\ref{tab:DSMCFLOW} defines the common flow state and sampling convention.

\subsection{Data Analysis Framework}

Dynamic Mode Decomposition (DMD) is used here to convert stored DSMC macrofield snapshots into modes with both spatial structure and time behavior~\cite{SCHMID_2010}. Related DSMC applications of data-driven modal analysis include unsteady high-speed ramp flows~\cite{Karpuzcuspod}. Let \(\mathbf{q}_k\) be the flattened DSMC macrofield at stored snapshot \(k\). For the DMD analysis below, \(\mathbf{q}_k\) is taken from the unforced second-mode dataset on the cropped \(1501\times63\) domain covering \(x=40\)--100 mm, and the operator is built from the last \(M=1{,}000\) snapshots. DMD fits a low-rank linear map that advances the data by one stored-snapshot interval,
\begin{equation}
    \mathbf{q}_{k+1} \approx \mathbf{A}\mathbf{q}_k .
\end{equation}

The numerical construction is summarized compactly as
\begin{equation}
    \begin{gathered}
    \mathbf{X}=[\mathbf{q}_1\ \mathbf{q}_2\ \cdots\ \mathbf{q}_{M-1}],\qquad
    \mathbf{Y}=[\mathbf{q}_2\ \mathbf{q}_3\ \cdots\ \mathbf{q}_{M}],\\
    \mathbf{X}=\mathbf{U}\boldsymbol{\Sigma}\mathbf{V}^*,\qquad
    \tilde{\mathbf{A}}=\mathbf{U}^*\mathbf{Y}\mathbf{V}\boldsymbol{\Sigma}^{-1},\qquad
    \tilde{\mathbf{A}}\mathbf{W}=\mathbf{W}\boldsymbol{\Lambda}.
    \end{gathered}
    \label{eqn:DMDcompact}
\end{equation}
Here \(\mathbf{X}\) and \(\mathbf{Y}\) are the time-shifted snapshot matrices, \(\tilde{\mathbf{A}}\) is the reduced advance operator, and the retained rank is \(r=30\). The full DMD modes are recovered as \(\boldsymbol{\Phi}=\mathbf{U}\mathbf{W}\), so each column \(\boldsymbol{\phi}_j\) is the spatial structure associated with eigenvalue \(\lambda_j\). The fixed rank limits noise-sensitive higher-rank content; rank sensitivity is treated as a limitation rather than as a uniqueness claim.

The key step is the conversion from the discrete DMD eigenvalue to continuous-time behavior. Writing \(\lambda_j=|\lambda_j|e^{i\theta_j}\),
\begin{align}
\Omega_j &= \frac{\ln(\lambda_j)}{\Delta t_{\mathrm{snap}}}
       = \sigma_j + i\omega_j, \label{eqn:DMDfrequency}\\
\sigma_j &= \frac{\ln|\lambda_j|}{\Delta t_{\mathrm{snap}}}, \quad
\omega_j = \frac{\theta_j}{\Delta t_{\mathrm{snap}}}.
\label{eqn:DMDgrowthfrequency}
\end{align}
The angular frequency \(\omega_j\) appears in the time exponential, while the dimensional frequency reported in the figures is \(\omega_j^*=\omega_j/(2\pi)\), in Hz or kHz. For this dataset, \(\Delta t_{\mathrm{snap}}=0.1\,\mu\mathrm{s}\), corresponding to a 10 MHz sampling frequency and a 5 MHz Nyquist limit, well above the \(\omega^*=100\)--500 kHz second-mode band targeted here. Thus each DMD output is a pair \((\boldsymbol{\phi}_j,\Omega_j)\): the mode gives the spatial disturbance structure, while the eigenvalue gives the frequency and temporal growth or decay. This eigenvalue-to-time-behavior map is the dynamic content of Dynamic Mode Decomposition and is the reason DMD is useful here as more than a mode-shape decomposition.

The extracted DMD modes can therefore be compared with the quantities used in classical Linear Stability Theory (LST). For example, the streamwise-velocity component of the data-derived modes can be written as
\begin{equation}
    u'(x, y, t) = \text{Re}\left( \sum_{j} \hat u_j(x,y)\, e^{i\,\omega_j t} \right),
    \label{eq:normal_modes_general_sum}
\end{equation}
where \(\hat u_j=[\boldsymbol{\phi}_j]_u\). Equation~\eqref{eq:normal_modes_general_sum} uses the same \(\omega_j\) defined in Eq.~\eqref{eqn:DMDgrowthfrequency}; the sign convention is immaterial because the LST comparison below uses the positive dimensional frequency \(|\omega_j|/(2\pi)\). The spatial part is read directly from \(\hat u_j\). Along the boundary-layer trajectory \(y=y_e(x)\), the phase and envelope are \(\theta_j(x)=\arg[\hat u_j(x,y_e(x))]\) and \(a_j(x)=|\hat u_j(x,y_e(x))|\). These quantities provide the mode location, wall-normal structure, wavelength, phase speed, and spatial amplification measures used below. In this workflow, LST is an a posteriori comparison for the DSMC-extracted modes, not an input that prescribes their frequency, shape, or location.

\section{Results}

\subsection{Mean Boundary-Layer State and Similarity Reference}

Before interpreting the unsteady spectra, the time-mean boundary layer must first be checked against the expected hypersonic laminar mean state. The reference is the zero-pressure-gradient compressible Blasius equations under the Howarth--Dorodnitsyn transformation. Let \(f'(\eta)=u/U_e\), \(g(\eta)=T/T_e\), and
\begin{equation}
\label{eq:chapman_rubesin_parameter}
C=\frac{\rho\mu}{\rho_e\mu_e}=g^{\sigma-1},
\end{equation}
where \(\sigma\) is the viscosity power-law exponent, and the subscript \(e\) denotes the local edge value outside the boundary layer. The similarity system used here follows the Crocco compressible-boundary-layer formulation~\cite{CroccoBL} and is
\begin{align}
\left(C f''\right)' + f f'' &= 0,\\
\left(C g'\right)' + Pr\, f g' + Pr\,C(\gamma-1)M_e^2\left(f''\right)^2 &= 0,
\end{align}
with \(f(0)=0\), \(f'(0)=0\), \(g(0)=T_w/T_e\), and \(f'(\infty)=g(\infty)=1\). This same solution also provides a generalized inflection-point diagnostic. For an ideal gas, \(\rho/\rho_e=1/g\), so the compressible inflection condition can be written as
\begin{equation}
\label{eq:generalized_inflection_condition}
\frac{d}{dy}\left(\rho\frac{dU}{dy}\right)=0
\quad\Longleftrightarrow\quad
\frac{d}{d\eta}\left(\frac{f''}{g^2}\right)=0.
\end{equation}
For repeatability, the CBL profiles use the same nominal edge state at every location: \(U_e=858.5\) m/s, \(T_e=52\) K, \(M_e=5.840\), \(N_e=1.2\times10^{24}\) m\(^{-3}\), \(\rho_e=5.5821\times10^{-2}\) kg/m\(^{3}\), \(T_w=300\) K, \(\gamma=1.4\), and \(Pr=0.72\). The viscosity is evaluated as \(\mu=\mu_{\mathrm{ref}}(T/T_{\mathrm{ref}})^\sigma\), with \(\mu_{\mathrm{ref}}=1.458731\times10^{-5}\) Pa\,s, \(T_{\mathrm{ref}}=273.15\) K, and \(\sigma=0.75\), giving \(\nu_e=7.5315\times10^{-5}\) m\(^2\)/s. The coarse DSMC and 5B DSMC mean profiles are compared using the stored record defined in Sec.~II at the sampled locations \(x_s=30\) and \(100\) mm.

\begin{figure}[H]
    \centering
    \includegraphics[width=0.92\textwidth]{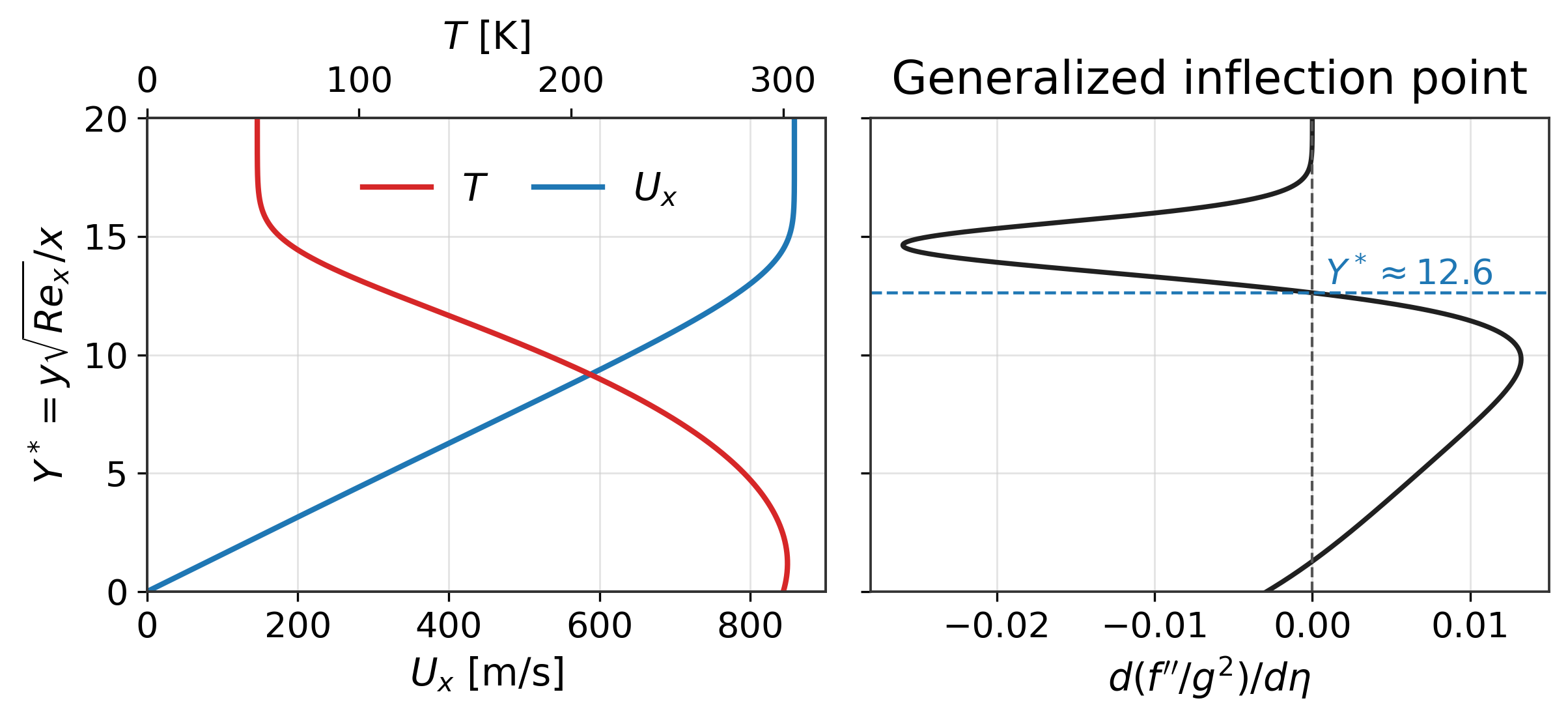}
    \caption{Compressible Blasius-type boundary-layer reference for the primary 5B edge state. The left panel shows \(T\) and \(U_x\) versus the self-similar height \(Y^*=y\sqrt{Re_x}/x\). The right panel shows the generalized-inflection diagnostic \(d(f''/g^2)/d\eta\), with the outer zero crossing at \(Y_i^*\approx\)12.6.}
    \label{fig:similarityref}
\end{figure}

\begin{figure}[H]
    \centering
    \includegraphics[width=0.95\textwidth,height=0.72\textheight,keepaspectratio]{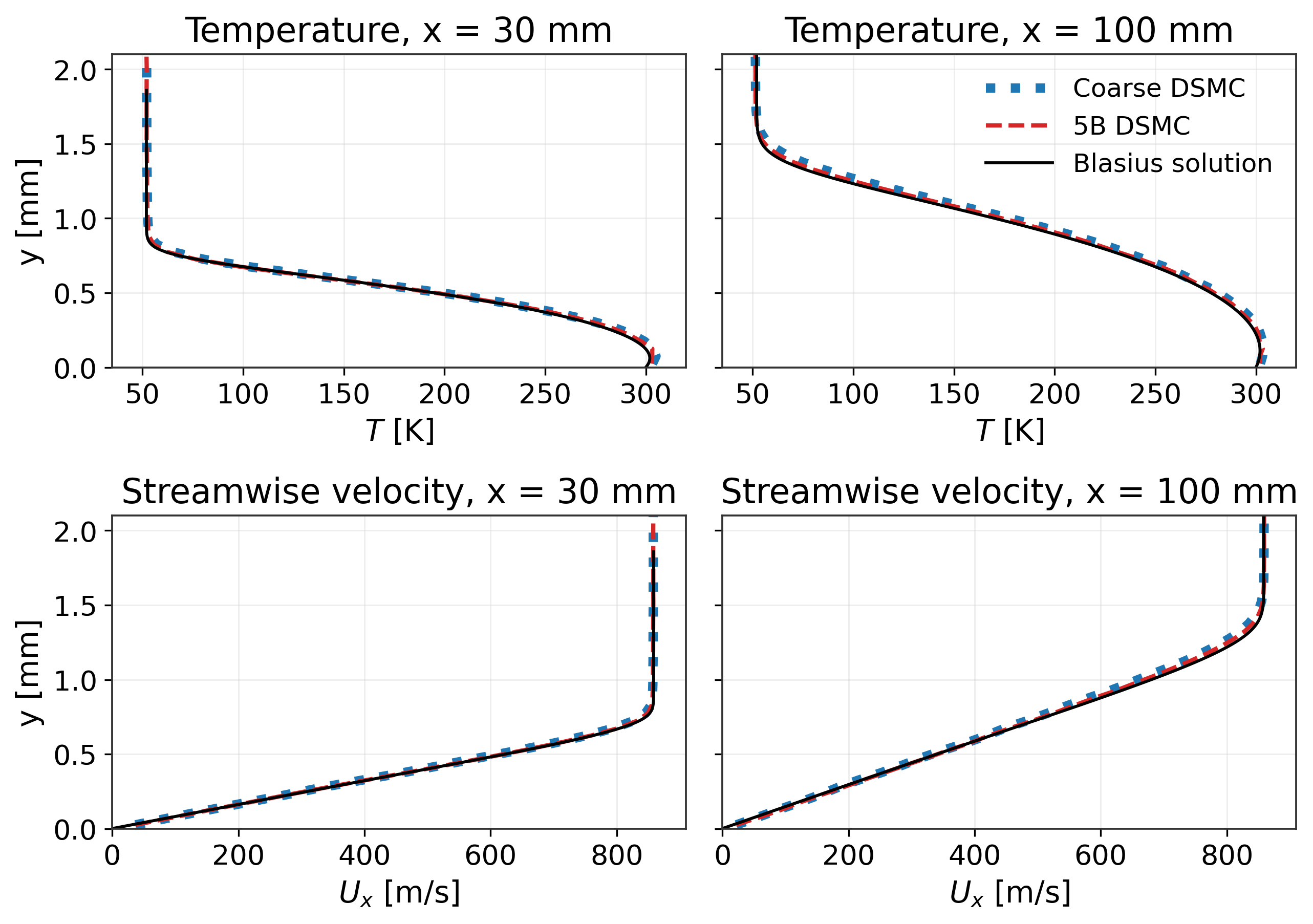}
    \caption{Coarse DSMC and primary 5B mean temperature and streamwise-velocity profiles at \(x=30\) and \(100\) mm compared with the compressible Blasius-type reference. Both DSMC profiles use the same stored record. The coarse DSMC case has roughly four times fewer particles than the 5B case but still recovers the same mean boundary-layer solution.}
    \label{fig:meanprofiles30100}
\end{figure}

Figure~\ref{fig:similarityref} first shows the compressible Blasius-type reference. The similarity temperature and streamwise-velocity profiles are plotted against \(Y^*=y\sqrt{Re_x}/x\), and the corresponding generalized-inflection diagnostic is shown on the same wall-normal coordinate. With the fixed nominal edge state, the outer generalized-inflection layer is \(Y_i^*=\)12.6. The DSMC mean profiles are checked for both the coarse DSMC and primary 5B cases. The particle counts and particles per \(\lambda_\infty^2\) for these two cases are given in Table~\ref{tab:particlecountcompare}. Figure~\ref{fig:meanprofiles30100} compares the coarse DSMC and 5B temperature and streamwise-velocity profiles with the compressible Blasius-type reference at \(x=30\) and \(100\) mm. The agreement shows that even the lower-particle-count DSMC case recovers the expected laminar mean boundary-layer solution. Since the primary 5B case has a larger total simulation particle count and a larger particles-per-\(\lambda_\infty^2\) value, the particle-count change mainly affects the fluctuation floor rather than the mean state.

\subsection{Flow Unsteadiness Analysis}

\subsubsection{RMS Definition and Fluctuation Diagnostics}
Throughout the remainder of this 120 mm analysis, RMS denotes the fluctuation amplitude over the stored record defined in Sec.~II. At any fixed spatial sampling location, let \(q\) denote any sampled DSMC variable of interest, such as pressure \(p\), density \(\rho\), streamwise velocity \(u\), wall-normal velocity \(v\), or temperature \(T\). Let \(q_t\) be the sampled value at stored time index \(t=1,\ldots,N_t\), with \(N_t=1{,}001\) for the RMS calculations below. The local mean and RMS fluctuation level are
\begin{equation}
\label{eq:rms_definition}
\bar{q}=\frac{1}{N_t}\sum_{t=1}^{N_t}q_t,\qquad
q'_{\mathrm{rms}}=\left[\frac{1}{N_t}\sum_{t=1}^{N_t}\left(q_t-\bar q\right)^2\right]^{1/2}.
\end{equation}
The prime denotes the mean-subtracted perturbation \(q'_t=q_t-\bar q\). Thus \(p'\), \(n'\), \(u'\), and \(T'\) refer to fluctuations about the local time mean, while \(\bar q\) is retained as the local reference value and as the normalization used for percentages.
The RMS maps and point diagnostics are built by applying this operation pointwise over the DSMC field, one spatial location at a time, for whichever variable \(q\) is under consideration.

Tables~\ref{tab:broadbandprobe_free} and~\ref{tab:filteredprobe_free} use the coarse and 5B DSMC cases defined in Table~\ref{tab:particlecountcompare}, with the same stored record and sampling interval. Percent RMS values are normalized by the corresponding local mean:
\begin{equation}
\label{eq:rms_percent_definition}
\%\left(q'_{\mathrm{rms}}\right) = 100\,\frac{q'_{\mathrm{rms}}}{\bar{q}}.
\end{equation}
For Tables~\ref{tab:broadbandprobe_free} and~\ref{tab:filteredprobe_free}, \(\bar{q}\) denotes the local time-averaged flow value at the sampled location. The 50 ns storage interval gives a resolved point-sampling band from about 20 kHz to the 10 MHz Nyquist limit. Table~\ref{tab:broadbandprobe_free} uses the freestream sample at \((x,y)=(1,2)\) mm; the local post-processing spacing is \(\Delta x \approx \Delta y \approx 0.05\) mm, so the DSMC measurement corresponds to one \(0.05\times0.05\) mm sampling cell over the full 20 kHz to 10 MHz band.
\begingroup
\setstretch{1.0}
\begin{table}[H]
\centering
\small
\setlength{\tabcolsep}{5.5pt}
\caption{Broadband freestream RMS values at \((x,y)=(1,2)\) mm. The coarse DSMC and 5B entries use the same stored record. All RMS values are computed from mean-subtracted perturbations over the full resolved point-sampling bandwidth, approximately 20 kHz to 10 MHz. The mean column lists the local reference value; percentages are normalized by each case's corresponding local mean.}
\label{tab:broadbandprobe_free}
\begin{tabular}{lccc}
\toprule
Quantity & Mean \(q\) & Coarse RMS & 5B RMS \\
\midrule
\(p'_{\mathrm{rms}}\) [Pa] & 828.2 & 4.465 \textit{(0.54\%)} & 2.421 \textit{(0.29\%)} \\
\(\rho'_{\mathrm{rms}}\) [kg/m\(^3\)] & \(5.58\times10^{-2}\) & \(2.47\times10^{-4}\) \textit{(0.44\%)} & \(1.37\times10^{-4}\) \textit{(0.25\%)} \\
\(u'_{\mathrm{rms}}\) [m/s] & 860.57 & 0.553 \textit{(0.06\%)} & 0.290 \textit{(0.03\%)} \\
\(T'_{\mathrm{rms}}\) [K] & 50.00 & 0.150 \textit{(0.30\%)} & 0.083 \textit{(0.17\%)} \\
\bottomrule
\end{tabular}
\end{table}
\endgroup

The broadband RMS ratios provide a direct check on the finite-particle interpretation. Here \(FNUM\) is the DSMC simulation ratio, or particle weight, defined as the number of real molecules represented by one simulation particle. If the fluctuation floor is governed by particle sampling, the expected coarse-to-5B RMS ratio scales approximately as
\begin{equation}
\label{eq:fnum_rms_scaling}
    \sqrt{\frac{FNUM_{\mathrm{coarse}}}{FNUM_{\mathrm{5B}}}}
    =
    \sqrt{\frac{2.88\times10^{11}}{8.525\times10^{10}}}
    =1.84 .
\end{equation}
The measured broadband coarse-to-5B ratios are 1.84 for pressure, 1.80 for density, 1.91 for streamwise velocity, and 1.81 for temperature. This agreement provides the particle-count consistency check: the particle-count change alters the stochastic fluctuation level while leaving the mean boundary layer essentially unchanged.

Table~\ref{tab:broadbandprobe_free} uses the full dynamic range available from the DSMC stored sampling step and one-cell DSMC measurement. A practical pressure measurement would generally retain a much lower bandwidth than the 10 MHz DSMC Nyquist limit and would average over a sensing area of order millimeters rather than one \(0.05\times0.05\) mm sampling cell. To isolate the bandwidth effect, Table~\ref{tab:filteredprobe_free} applies a finite \(20\)--\(100\) kHz filter to the same freestream point \((x,y)=(1,2)\) mm while keeping the same DSMC measurement. This produces a practical finite-band DSMC noise level for the chosen particle count, not a physical probe-response or flight-instrumentation model. Let \(q'_t\) be the mean-subtracted trace of any variable \(q\), and let \(Q_m\) be its discrete Fourier coefficient at dimensional frequency \(\omega_m^*\). Define the band mask
\begin{equation}
\label{eq:bandlimited_rms}
\begin{aligned}
H_m &=
\begin{cases}
1, & 20~\mathrm{kHz}<\omega_m^*\le100~\mathrm{kHz},\\
0, & \mathrm{otherwise},
\end{cases}
\\
q'_{B,t} &= \mathcal{F}^{-1}\{H_m Q_m\}_t,\\
q'_{\mathrm{rms},B}
&= \left(\frac{1}{N_t}\sum_{t=1}^{N_t}\left(q'_{B,t}\right)^2\right)^{1/2}.
\end{aligned}
\end{equation}
Here \(B\) denotes the retained band. The mask \(H_m\) keeps the selected dimensional frequencies and sets all other Fourier coefficients to zero before transforming back to time. This same operation gives every entry in Table~\ref{tab:filteredprobe_free}. The selected point lies in the freestream above the plate boundary layer, so it measures the disturbance environment entering the boundary layer rather than the amplified boundary-layer response.

\begingroup
\setstretch{1.0}
\begin{table}[H]
\centering
\small
\setlength{\tabcolsep}{5.5pt}
\caption{Filtered freestream RMS values at \((x,y)=(1,2)\) mm, retaining only \(20\) kHz \(< \omega^* \le 100\) kHz. The pressure row gives the finite-band DSMC numerical noise metric for the one-cell DSMC measurement; the other rows are companion DSMC state diagnostics. The case windows, local means, and normalization convention are the same as in Table~\ref{tab:broadbandprobe_free}.}
\label{tab:filteredprobe_free}
\begin{tabular}{lccc}
\toprule
Quantity & Mean \(q\) & Coarse RMS & 5B RMS \\
\midrule
\(p'_{\mathrm{rms}}\) [Pa] & 828.2 & 0.568 \textit{(0.07\%)} & 0.284 \textit{(0.03\%)} \\
\(\rho'_{\mathrm{rms}}\) [kg/m\(^3\)] & \(5.58\times10^{-2}\) & \(3.60\times10^{-5}\) \textit{(0.06\%)} & \(1.47\times10^{-5}\) \textit{(0.03\%)} \\
\(u'_{\mathrm{rms}}\) [m/s] & 860.57 & 0.075 \textit{(0.01\%)} & 0.054 \textit{(0.01\%)} \\
\(T'_{\mathrm{rms}}\) [K] & 50.00 & 0.020 \textit{(0.04\%)} & 0.010 \textit{(0.02\%)} \\
\bottomrule
\end{tabular}
\end{table}
\endgroup

\noindent Tables~\ref{tab:broadbandprobe_free} and~\ref{tab:filteredprobe_free} show the same particle-count trend under two numerical measurement definitions. The broadband table uses the full resolved DSMC point-sampling band, while the filtered table restricts the same one-cell signal to a finite low-frequency band. Under this definition, the 5B pressure level is \(p'_{\mathrm{rms}}/\bar{p}=0.03\%\). This finite-band value is the practical DSMC fluctuation level for the chosen particle count and sampled point.

\subsection{Data-Derived Extraction of LST Modal Quantities}

The next question is whether the unseeded DSMC data contain the frequencies, wavenumbers, mode shapes, and amplification measures normally supplied by a stability calculation. Classical LST begins by assuming that a small disturbance to the mean boundary layer can be represented by a normal mode. For a disturbance vector \(\mathbf{q}'=[\rho',u',v',T',p']^T\), the local form is
\begin{equation}
    \mathbf{q}'(x,y,t)
    =
    \mathrm{Re}\left\{
    \hat{\mathbf{q}}(y)\,
    A(x)\,
    \exp\left[i\left(\alpha x-\omega t\right)\right]
    \right\},
    \label{eq:lstlocalansatz}
\end{equation}
where \(\omega\) is the temporal angular frequency, \(\alpha\) is the real streamwise wavenumber, \(\lambda_x=2\pi/\alpha\) is the wavelength, \(c_p=\omega/\alpha\) is the phase speed, \(\hat{\mathbf{q}}(y)\) is the wall-normal eigenfunction, and \(A(x)\) is the streamwise amplitude envelope. In a slowly varying boundary layer, the same idea is written with a spatially varying phase and envelope,
\begin{equation}
    \mathbf{q}'(x,y,t)
    =
    \mathrm{Re}\left\{
    \hat{\mathbf{q}}(y;x)\,
    A(x)\,
    \exp\left[i\left(\int_{x_0}^{x}\alpha(\xi)\,d\xi-\omega t\right)\right]
    \right\}.
    \label{eq:lstslowansatz}
\end{equation}
Equations~\eqref{eq:lstlocalansatz} and \eqref{eq:lstslowansatz} define the quantities used below for the DSMC-DMD and LST comparison: dimensional frequency \(\omega^*\), wavenumber, wavelength, phase speed, wall-normal structure, streamwise location and extent, and amplification or decay.

\begin{figure}[H]
    \centering
    \includegraphics[width=0.72\textwidth]{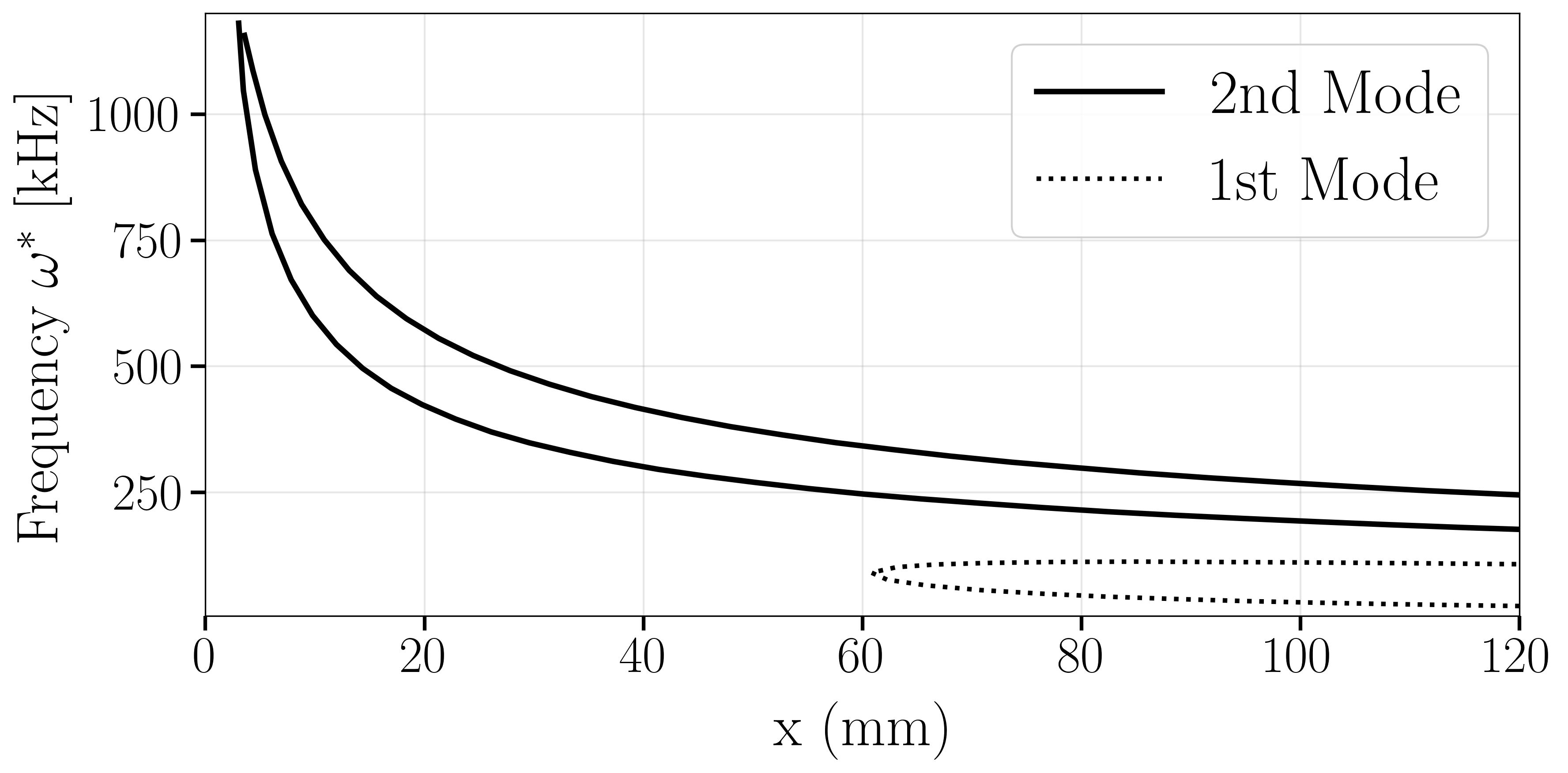}
    \caption{Neutral stability curves for the Mach 6 flat-plate reference, transformed from the Ozgen et al.~\cite{ozgenLinearStabilityAnalysis2008} results to dimensional frequency \(\omega^*\) and streamwise position using the present flow parameters. The second-mode neutral envelope identifies the \(\omega^*\)-location region where LST predicts spatial amplification; the first-mode envelope is shown for context.}
    \label{fig:lstneutralcurves}
\end{figure}

The LST reference used in this manuscript is shown in Fig.~\ref{fig:lstneutralcurves}. The neutral curves are based on the Mach 6 flat-plate stability results of Ozgen et al.~\cite{ozgenLinearStabilityAnalysis2008}, re-expressed in dimensional frequency \(\omega^*\) and streamwise position using the present edge-state parameters. The enclosed second-mode region gives the expected \(\omega^*\)-location window for spatial amplification. It is used only after the DSMC-DMD extraction, so the stability calculation provides an external reference rather than a prescribed input to the data analysis.

DMD estimates analogous modal characteristics from the DSMC data without imposing this ansatz in advance. The fluctuating DSMC macrofield is approximated as
\begin{equation}
    \mathbf{q}'(x,y,t_k)
    \approx
    \mathrm{Re}\left\{
    \sum_j b_j\,\boldsymbol{\phi}_j(x,y)\,\exp(\Omega_j t_k)
    \right\},
    \qquad
    \Omega_j=\frac{\ln(\lambda_j)}{\Delta t_{\mathrm{snap}}}
    =\sigma_j+i\omega_j .
    \label{eq:dmdmodalexpansion}
\end{equation}
Here \(\lambda_j\) is the discrete DMD eigenvalue, \(\Omega_j\) is the continuous-time eigenvalue, and \(\boldsymbol{\phi}_j(x,y)\) is the complex spatial mode. A mode can then be decomposed into the same ingredients that appear in Eq.~\eqref{eq:lstslowansatz}:
\begin{equation}
    \boldsymbol{\phi}_j(x,y)
    =
    \hat{\mathbf{q}}_j(y;x)\,
    \mathcal{A}_j(x)\,
    \exp[i\theta_j(x)] .
    \label{eq:dmdspatialfactor}
\end{equation}
The imaginary part of \(\Omega_j\) gives the dimensional frequency \(\omega_j^*=\omega_j/(2\pi)\), while the real part gives the temporal growth or decay rate of that extracted component over the sampled window. The spatial phase \(\theta_j(x)\) gives the local wavenumber through \(\alpha_j=d\theta_j/dx\), or equivalently through the dominant Fourier peak of a streamwise trace through the mode. The magnitude \(\mathcal{A}_j(x)=|\boldsymbol{\phi}_j|\) gives the envelope and \(d\ln \mathcal{A}_j/dx\) gives local spatial growth or decay. The wall-normal variation of \(\boldsymbol{\phi}_j\) gives the eigenfunction-like structure. Thus DMD does not merely identify \(\omega^*\). It estimates the modal frequency, temporal growth or decay, spatial phase, amplitude envelope, and wall-normal structure from the unsteady DSMC data.

Koopman/DMD-based analysis has also been used to extract second-mode wavepacket structure from hypersonic schlieren data~\cite{Ghannadian2024}; here DMD is applied to DSMC macrofield snapshots rather than schlieren intensity fields. The DMD calculation returns modes at different dimensional frequencies and at different locations in the boundary layer. Each retained mode has a discrete eigenvalue, a continuous-time behavior through \(\Omega_j\), and a complex spatial field \(\boldsymbol{\phi}_j\).

\begin{figure}[H]
    \centering
    \begin{subfigure}[b]{0.98\textwidth}
        \includegraphics[width=\linewidth]{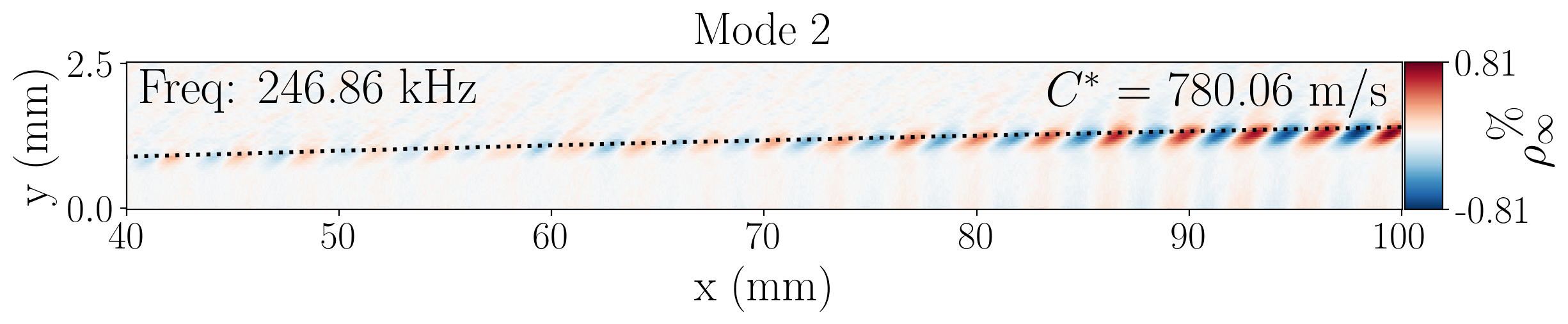}
        \caption{Density perturbation mode at \(\omega^*=247\) kHz with the boundary-layer trajectory shown by the dotted curve.}
        \label{subfig:dmdmode247finaltwo}
    \end{subfigure}\\[4pt]
    \begin{subfigure}[b]{0.72\textwidth}
        \includegraphics[width=\linewidth]{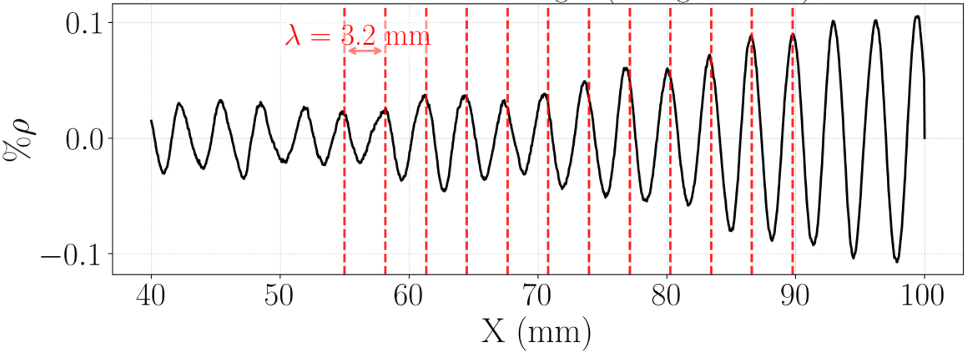}
        \caption{Spatial oscillation extracted along the boundary-layer trajectory.}
        \label{subfig:dmdprobe247finaltwo}
    \end{subfigure}\\[4pt]
    \begin{subfigure}[b]{0.72\textwidth}
        \includegraphics[width=\linewidth]{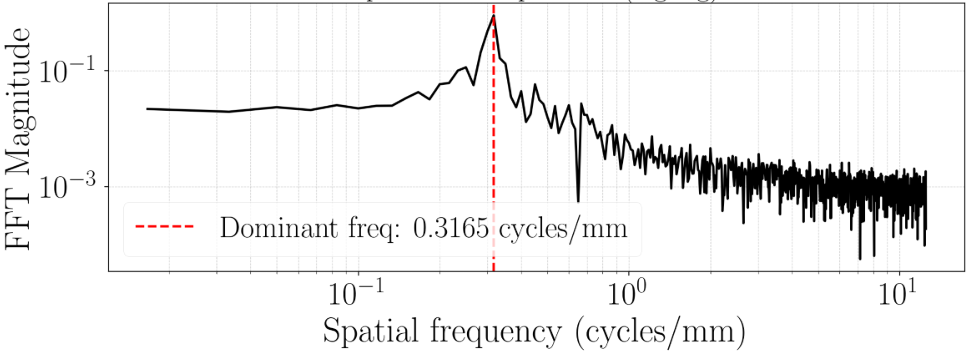}
        \caption{Fourier spectrum used to identify the dominant streamwise spatial frequency.}
        \label{subfig:dmdfft247finaltwo}
    \end{subfigure}
    \caption{Extraction of traveling-wave quantities from one DMD mode. The DMD eigenvalue gives \(\omega_j^*=246.86\) kHz, the modal field shows where the mode is located and gives its wall-normal structure, the streamwise trace gives \(\kappa_{x,j}=0.3165\) cycles/mm and \(\lambda_{x,j}=3.16\) mm, and the product \(\omega_j^*\lambda_{x,j}\) gives \(c_{p,j}\approx780\) m/s.}
    \label{fig:dmdwavelengthfinaltwo}
\end{figure}

Figure~\ref{fig:dmdwavelengthfinaltwo} shows how the \(\omega^*=247\) kHz DMD mode provides the traveling-wave quantities that appear in LST. Figure~\ref{subfig:dmdmode247finaltwo} is the real part of the DMD spatial mode, \(\mathrm{Re}[\boldsymbol{\phi}_j(x,y)]\), normalized by \(\rho_\infty\). This subfigure shows where the mode is located, its wall-normal structure, and its phase pattern. The dimensional frequency \(\omega_j^*=246.86\) kHz is obtained directly from the DMD eigenvalue through Eqs.~\eqref{eqn:DMDfrequency}--\eqref{eqn:DMDgrowthfrequency}; it is not imposed by a stability calculation. The dotted curve is the boundary-layer trajectory \(y_e(x)\), along which the one-dimensional modal trace is extracted,
\begin{equation}
    s_j(x)=\mathrm{Re}\left[\boldsymbol{\phi}_j(x,y_e(x))\right],
\end{equation}
as shown in Fig.~\ref{subfig:dmdprobe247finaltwo}. Figure~\ref{subfig:dmdfft247finaltwo} is the Fourier spectrum of this modal trace. If the spectrum has a dominant peak at \(\kappa_{x,j}\) cycles per unit length, then
\begin{equation}
    \lambda_{x,j}=\frac{1}{\kappa_{x,j}}, \qquad
    \alpha_j=2\pi\kappa_{x,j}, \qquad
    c_{p,j}=\frac{\omega_j}{\alpha_j}=\omega_j^*\lambda_{x,j}.
\end{equation}
For this mode, the dominant peak is \(\kappa_{x,j}=0.3165\) cycles/mm, giving \(\lambda_{x,j}=3.16\) mm, \(\alpha_j=1.99\) rad/mm, and \(c_{p,j}=\omega_j^*\lambda_{x,j}\approx780\) m/s. These are the dimensional frequency \(\omega^*\), wavenumber, wavelength, and phase speed that appear in the LST normal-mode description, but here they are measured directly from the DSMC snapshots.

The remaining LST quantity is the streamwise amplification. In LST, spatial growth is commonly summarized by the \(N\)-factor,
\begin{equation}
    N(x,\omega^*)=\int_{x_0}^{x}-\alpha_i(\xi,\omega^*)\,d\xi,
    \qquad
    e^{N(x,\omega^*)}=\frac{A(x,\omega^*)}{A_0(\omega^*)} ,
    \label{eq:nfactorlstfinaltwo}
\end{equation}
where \(A_0\) is the initial unamplified amplitude and \(-\alpha_i\) is the spatial growth rate. The DMD analogue used here is obtained directly from the modal envelope. It should be read as an \(N\)-factor-like envelope, not as a true transition-prediction \(N\)-factor. Along the same boundary-layer trajectory,
\begin{equation}
    E_j(R)
    =
    \frac{\mathcal{A}_j(R)}{\mathcal{A}_j(R_0)}
    ,
    \qquad
    N_j(R)=\ln E_j(R).
    \label{eq:dmdnfactorfinaltwo}
\end{equation}
This envelope is a spatial amplitude measure extracted from \(\boldsymbol{\phi}_j\); it is distinct from the temporal growth rate \(\sigma_j\) encoded in \(\Omega_j\).
For the representative mode in Fig.~\ref{fig:dmdwavelengthfinaltwo}, the increasing amplitude of the signed oscillation along \(y_e(x)\) is the visible trace of this envelope; \(E_j\) is computed from \(\mathcal{A}_j(x)=|\boldsymbol{\phi}_j|\), not from the signed real-part signal. The envelope can therefore be used as an \(N\)-factor-like measure for a retained DMD mode without importing an LST growth rate. The comparison with LST remains an a posteriori check: the frequency, wavelength, phase speed, location, structure, and modal envelope are obtained from DMD first, and LST is used afterward to test whether these DSMC-derived characteristics occupy the expected second-mode range.

\section{Conclusions}

This work demonstrates a DSMC-based procedure for studying hypersonic boundary-layer unsteadiness without prescribing an inlet wave, wall forcing, modal shape, or disturbance spectrum. The Mach 6 flat-plate simulation contains only the finite-particle stochastic fluctuation field carried by the particle representation. The mean-flow comparison shows that both the coarse DSMC and primary 5B cases recover the same laminar compressible Blasius-type boundary layer at \(x=30\) and \(100\) mm, even though the coarse case uses roughly four times fewer particles. The particle-count change therefore mainly changes the fluctuation level rather than the mean state.

The RMS analysis then gives the simulation disturbance level under explicitly stated measurement definitions. For the freestream sample location, restricting the one-cell signal to the finite \(20\)--\(100\) kHz band gives a 5B pressure fluctuation level of \(p'_{\mathrm{rms}}/\bar{p}=0.03\%\). This is the practical finite-band DSMC fluctuation level for the chosen particle count and sampled point. The DMD analysis shows that unseeded DSMC snapshots can be reduced to traveling-wave quantities expressed in the variables used by classical stability analysis. The representative density mode demonstrates that the DMD mode and eigenvalue together contain the modal frequency, wavelength, phase speed, wall-normal structure, streamwise location, temporal behavior, and an \(N\)-factor-like amplitude envelope. These quantities are obtained from the DSMC data rather than supplied as inputs. LST is then used after the extraction to test whether these DSMC-derived characteristics are consistent with the expected second-mode response.

Together, we claim that DSMC can be used not only as a mean-flow solver, but to study receptivity and flow instabilities. In the present Mach 6 flat-plate case, the second-mode disturbance field develops from the stochastic fluctuation field carried by the DSMC particle representation rather than from an external a priori seeded excitation. Once the unsteady DSMC macrofields are stored, data-driven modal analysis can extract the instability characteristics directly from the stored unsteady DSMC macrofield data: for each retained mode, the dimensional frequency, temporal growth or decay, wavelength, phase speed, mode location, mode shape, and amplitude envelope. The comparison with LST therefore shows that DSMC-DMD can identify second-mode wave quantities from the simulation data alone, and it defines a procedure for studying other hypersonic flow instabilities whose modal structure is not known in advance.

\section*{Acknowledgments}
This work was supported by the Office of Naval Research under Grant No. N00014-23-1-2839, "Kinetic Treatment of Sources and Mechanisms that Drive Unsteady, Shock-dominated Flow Instability" (Program Officer: Dr. Eric Marineau). Computational resources were provided by the Texas Advanced Computing Center (TACC) at the University of Texas through the Frontera supercomputer (Project No. CTS23002), funded by the National Science Foundation (NSF), and by Purdue University's Anvil supercomputer through ACCESS allocation PHY250034.

\newpage
\bibliography{sample}

\end{document}